\documentclass[
 aip, jmp
 reprint,
 amsmath,amssymb,
]{revtex4-1}

\usepackage{graphicx}
\usepackage{dcolumn}
\usepackage{bm}

\begin{document}

\title{An improved algorithm for reconstructing reflectionless potentials}
\author{Matti Selg}
\affiliation{Institute of Physics of the University of Tartu, Ravila 14c,
50411, Tartu, Estonia}
\date{\today}

\begin{abstract}
A fully algebraic approach to reconstructing one-dimensional reflectionless
potentials is described. A simple and easily applicable general formula is
derived, using the methods of the theory of determinants. In particular,
useful properties of special determinants -- the alternants -- have been
exploited. The main formula takes an especially simple form if one aims to
reconstruct a symmetric reflectionless potential. Several examples are
presented to illustrate the efficiency of the method.

\end{abstract}

\maketitle

\section{Introduction}

Formulation and solution of inverse problems is an increasingly important
field of scientific research. However, compared with a well-posed (in
Hadamard's sense) direct or forward problem, the corresponding inverse
problem is much more difficult and, as a rule, ill-posed. Inverse scattering
problem can be considered an exception to this rule. Namely, in the simplest
one-dimensional case, the inverse scattering theory provides strict
mathematical criteria for the existence, uniqueness and stability of the
solution. It means that in this particular case the inverse problem is
well-posed as well.

The related forward problem is the solution of the simplest time-independent
Schr\"{o}dinger equation%
\begin{equation}
\Psi ^{\prime \prime }(x)=\frac{V(x)-E}{C}\Psi (x),~C\equiv \frac{\hbar ^{2}%
}{2m}  \label{v1}
\end{equation}%
for a given potential $V(x)$, and subjected to appropriate physical boundary
conditions. Eq. (\ref{v1}) can be easily solved numerically and thus, in
principle, all spectral characteristics of the potential $V(x)$ can be
accurately ascertained.

The inverse problem is to determine the \textbf{unknown} potential starting
from the known spectral characteristics. This is a serious task even in this
simple case for the following reasons:

\begin{itemize}
\item[1.] It is not obvious what kind of input information is actually
needed to solve the problem uniquely.

\item[2.] There must be a theoretical basis (a fundamental equation) which
enables to solve the problem.

\item[3.] Apart from the theoretical difficulties, another important
question arises: how to obtain the necessary input data?

\item[4.] Even if the mentioned principal barriers could be overcome, the
computational-technical solution of the problem is not at all trivial.
\end{itemize}

Issues 1 and 2 have been successfully resolved in early 1950s for the class
of potentials on the half line: $x\in \left[ 0,\infty \right) $. The
necessary and sufficient conditions for the unique solution of the inverse
problem have been formulated in a series of outstanding theoretical works by
Marchenko \cite{Marchenko1, Marchenko2}, Gelfand, Levitan \cite{GL}, Krein 
\cite{Krein1, Krein2} and others (see, e.g., \cite{CS}, section III.7 for a
brief overview). In addition, three different methods to solve the problem
have been worked out, based on the integral equations by Gelfand-Levitan 
\cite{GL}, Marchenko \cite{Marchenko2} and Krein \cite{Krein2}.

Solution of the inverse scattering problem on the full line ($-\infty
<x<\infty $) is a more challenging problem, which was first addressed by Kay 
\cite{Kay}, Moses \cite{Moses}, and few years later in a series of papers by
Faddeev \cite{Faddeev58, Faddeev59, Faddeev64, Faddeev74}. These fundamental
studies provide full description of the solution procedure, while the
correct necessary and sufficient criteria for the uniqueness of the solution
were given by Marchenko (see \cite{MarchenkoBook}, section III.5). These
criteria apply to the following pair of integral equations \cite{Moses,
Faddeev74, MarchenkoBook}:%
\begin{equation}
K_{1}(x,y)+A_{1}(x+y)+\int\limits_{x}^{\infty }A_{1}(z+y)K_{1}(x,z)dz=0,
\label{v2}
\end{equation}%
\begin{equation}
K_{2}(x,y)+A_{2}(x+y)+\int\limits_{-\infty }^{x}A_{2}(z+y)K_{2}(x,z)dz=0,
\label{v3}
\end{equation}%
which are formally very similar to the Marchenko equation on the half line
(see \cite{MarchenkoBook}, p. 218). On the other hand, their
operator-theoretical content is closer to the Gelfand-Levitan approach.
Therefore, as a kind of compromise, Eqs. (\ref{v2})-(\ref{v3}) are often called
Gelfand-Levitan-Marchenko (GLM)\ equations. The kernel $A_{1}$ in Eq. (\ref%
{v2}) (where $y>x$)\ is completely specified by the so-called right
scattering data, while the kernel $A_{2}$ in Eq. (\ref{v3}) ($y<x$) is
determined by the left scattering data. These two sets of input data are
equivalent: the left data are uniquely determined by the right ones and 
\textit{vice versa}. In the following analysis we will rely on Eq. (\ref{v2}%
). If one is able to solve this integral equation then the potential is
given by%
\begin{equation}
V(x)=-2C\frac{dK_{1}(x,x)}{dx}.  \label{v4}
\end{equation}

Now, let us briefly discuss Point 3 in the above list. Unfortunately, the
criteria for the uniqueness of the solution are so strict that it is nearly
impossible to get all the necessary input data experimentally. However, if
one sets an additional constraint that the resulting potential must be
reflectionless then the inverse scattering problem can be solved much more
easily. Moreover, a \textbf{symmetric} reflectionless potential is uniquely
determined if its full spectrum of bound states is known. In the following
analysis it is assumed that the potential $V(x)$ that corresponds to Eq. (%
\ref{v4}) is reflectionless by definition.

As the general principles of building confining reflectionless potentials
are long known \cite{Faddeev74, Thacker}, a natural question arises: is
there any need to revisit the topic? A motivation comes from Point 4 stated
above: if the total number of bound states is large then the technical side
of the procedure becomes important. This in turn motivates development of
more efficient algorithms. In this paper, a new analytic algorithm is
derived, which enables to easily calculate a reflectionless potential with
an arbitrary number ($N$) of bound states.

The paper is organized as follows. In Sec. II, the general principles are
briefly described, which form the overall basis for the approach. \ Secs.
III and IV make an excursion to the theory of determinants, the benefits of
which are described and illustrated in Sec. V. Finally, Sec. VI concludes
the work.

\section{Reconstruction of reflectionlesss potentials: universal recipe}

Suppose we are given $2N$ parameters for an unknown reflectionless potential 
$V(x)$: the positions of $N$ discrete energy levels $E_{n}=-C\kappa _{n}^{2}$
and $N$ norming constants$\ C_{n}$ ($n=1,2,...,N$) for the Jost solution $%
\Psi _{1}(i\kappa _{n},x)\ $of Eq. (\ref{v1}), so that $\Psi _{1}(i\kappa
_{n},x)\rightarrow \exp (-\kappa _{n}x)$ as $x\rightarrow +\infty ,$ and%
\begin{equation}
\int\limits_{-\infty }^{\infty }\Psi _{n}^{2}(x)dx=1,\text{ }\Psi _{n}\equiv
C_{n}\Psi _{1}(i\kappa _{n},x).  \label{v5}
\end{equation}%
Then it can be shown \cite{Thacker} that%
\begin{equation}
\Psi _{n}(x)=-\dfrac{1}{\Lambda _{n}(x)}\dfrac{\det \left[ A^{(n)}\right] }{%
\det \left( A\right) }.  \label{v6}
\end{equation}%
Here $A$ is a symmetric matrix with the following elements:%
\begin{equation}
A_{mn}=\delta _{mn}+\dfrac{\Lambda _{m}(x)\Lambda _{n}(x)}{\kappa
_{m}+\kappa _{n}},  \label{v7}
\end{equation}%
$A^{(n)}$ is obtained from $A$ by replacing the $n$th column with its
derivative, and%
\begin{equation}
\Lambda _{n}(x)\equiv C_{n}\exp (-\kappa _{n}\,x).  \label{v8}
\end{equation}

A simple formula can also be obtained for the potential \cite{Faddeev74,
Thacker}:%
\begin{equation}
V(x)=-2C\dfrac{d^{2}}{dx^{2}}\left\{ \ln \left[ \det (A)\right] \right\} ,
\label{v9}
\end{equation}%
which is uniquely determined by parameters $\kappa _{n}$ and $C_{n}.$

Note that the potential remains unchanged if $\det \left( A\right) $ is
multiplied by a function $\exp \left( \alpha x+\beta \right) $, where $%
\alpha $ and $\beta $\ are arbitrary constants. Consequently, we can
multiply, for example, any row of the initial matrix by a function $\exp %
\left[ \kappa _{n}(x-x_{n})\right] $ ($n=1,2,...,N$), where $x_{n}$ are new
parameters, equivalent to norming constants: 
\begin{equation}
\exp \left( 2\kappa _{n}x_{n}\right) \equiv \dfrac{C_{n}^{2}}{2\kappa _{n}},
\label{v10}
\end{equation}%
and thus, according to definition (\ref{v8}), 
\begin{equation}
\dfrac{\Lambda _{n}^{2}(x)}{2\kappa _{n}}=\exp \left[ -2\kappa _{n}(x-x_{n})%
\right] .  \label{v11}
\end{equation}%
As a result, we get another matrix which contains full information for
reconstructing the potential according to Eq. (\ref{v9}):%
\begin{equation}
\begin{array}{c}
\tilde{A}_{N}=\tilde{B}_{N}+\tilde{C}_{N}, \\ 
V(x)=-2C\dfrac{d^{2}}{dx^{2}}\left\{ \ln \left[ \det (\tilde{A}_{N})\right]
\right\} .%
\end{array}
\label{v12}
\end{equation}%
Here%
\begin{equation}
\tilde{B}_{N}\equiv \left( 
\begin{array}{cccc}
e^{\kappa _{1}(x-x_{1})} & 0 & ... & 0 \\ 
0 & e^{\kappa _{2}(x-x_{2})} & ... & 0 \\ 
0 & ... & ... & ... \\ 
0 & ... & ... & e^{\kappa _{N}(x-x_{N})}%
\end{array}%
\right)  \label{v13}
\end{equation}%
and%
\begin{equation}
\tilde{C}_{N}\equiv \left( 
\begin{array}{cccc}
e^{-\kappa _{1}(x-x_{1})} & \frac{2\sqrt{\kappa _{1}\kappa _{2}}}{\kappa
_{1}+\kappa _{2}}e^{-\kappa _{2}(x-x_{2})} & ... & \frac{2\sqrt{\kappa
_{1}\kappa _{N}}}{\kappa _{1}+\kappa _{N}}e^{-\kappa _{n}(x-x_{N})} \\ 
\frac{2\sqrt{\kappa _{1}\kappa _{2}}}{\kappa _{1}+\kappa _{2}}e^{-\kappa
_{1}(x-x_{1})} & e^{-\kappa _{2}(x-x_{2})} & ... & \frac{2\sqrt{\kappa
_{2}\kappa _{N}}}{\kappa _{2}+\kappa _{N}}e^{-\kappa _{n}(x-x_{N})} \\ 
... & ... & ... & ... \\ 
\frac{2\sqrt{\kappa _{1}\kappa _{N}}}{\kappa _{1}+\kappa _{N}}e^{-\kappa
_{1}(x-x_{1})} & \frac{2\sqrt{\kappa _{2}\kappa _{N}}}{\kappa _{2}+\kappa
_{N}}e^{-\kappa _{2}(x-x_{2})} & ... & e^{-\kappa _{N}(x-x_{N})}%
\end{array}%
\right) .  \label{v14}
\end{equation}%
%
%
%
%

\textbf{Remark: }A subscript was added to denote the number of the bound
states (and the rank of the matrix).

From now on, the determinants having the structure $\det (A)\cdot \exp
\left( \alpha x+\beta \right) $ with the elements $A_{mn}$ defined in Eq. (%
\ref{v7}), will be sometimes called $\tau $-functions (as is common in
soliton theory). In the next two sections it will be shown that such
determinants can be easily calculated even for an arbitrarily large $N$.

\section{General formula for the $\protect\tau $-functions}

To evaluate a non-trivial determinant, one can use the Laplace expansion
(see \cite{Meyer}, p. 487) in terms of the fixed row (or column) indices.
For example, choosing a set of indices $m_{1},$ $m_{2},...,m_{k}$ for an
arbitrary $N\times N$-matrix $A_{N}$, so that $1\leq m_{1}<m_{2}<...\leq N$,
we get 
\begin{gather}
\det \left( A_{N}\right) =\sum\limits_{1\leq n_{1}<n_{2}<...\leq N}\det
A\left( m_{1}m_{2}...m_{k}\right\vert n_{1}n_{2}...n_{k})\times  \label{v15}
\\
\det \widehat{A}\left( m_{1}m_{2}...m_{k}\right\vert n_{1}n_{2}...n_{k}). 
\notag
\end{gather}%
Here $A\left( m_{1}m_{2}...m_{k}\right\vert n_{1}n_{2}...n_{k})$ is a $%
k\times k$-submatrix of $A_{N}$ that lies on the intersection of rows $%
m_{1}, $ $m_{2},...,m_{k}$ and columns $n_{1},$ $n_{2},...,n_{k}$, while 
\begin{gather*}
\det \widehat{A}\left( m_{1}m_{2}...m_{k}\right\vert n_{1}n_{2}...n_{k})= \\
(-1)^{m_{1}+...+m_{k}+n_{1}...+n_{k}}M\left( m_{1}m_{2}...m_{k}\right\vert
n_{1}n_{2}...n_{k}),
\end{gather*}%
and $M\left( m_{1}m_{2}...m_{k}\right\vert n_{1}n_{2}...n_{k})$ is a minor
obtained from $\det \left( A_{N}\right) $ by deleting rows $m_{1},$ $%
m_{2},...,m_{k}$ and columns $n_{1},$ $n_{2},...,n_{k}$.

Consequently, applying Eq. (\ref{v15}) to the $\tau $-function defined by
Eqs. (\ref{v12})-(\ref{v14}),%
\begin{gather}
\tau _{N}\equiv \det \left( \tilde{A}_{N}\right) =a_{0}\exp (\alpha
_{0})+\sum_{i=1}^{N}a_{i}\exp (\alpha _{i})+  \label{v16} \\
\sum_{1\leq i<j\leq N}a_{ij}\exp (\alpha _{ij})+\sum\limits_{1\leq i<j<k\leq
N}a_{ijk}\exp (\alpha _{ijk})+  \notag \\
...+a_{123...N}\exp (\alpha _{123...N}),  \notag
\end{gather}%
where the coefficients $a_{0},$ $a_{i},$ $a_{ij},$ $a_{ijk}$,... as well as
the corresponding arguments of the exponents can be easily fixed with the
help of Eqs. (\ref{v12})-(\ref{v14}). Indeed,%
\begin{gather}
\alpha _{0}=\sum_{l=1}^{N}\kappa _{l}(x-x_{l}),\ \alpha
_{i}=\sum_{l=1}^{N}(-1)^{\delta _{il}}\kappa _{l}(x-x_{l}),  \label{v17} \\
\alpha _{ij}=\sum_{m=1}^{N}(-1)^{\delta _{im}+\delta _{jm}}\kappa
_{m}(x-x_{m}),  \notag \\
\alpha _{ijk}=\sum\limits_{n=1}^{N}(-1)^{\delta _{in}+\delta _{jn}+\delta
_{kn}}\kappa _{n}(x-x_{n}),  \notag \\
...\alpha _{123...N}=\sum\limits_{n=1}^{N}(-1)^{\delta _{1n}+\delta
_{2n}+...\delta _{Nn}}\kappa _{n}(x-x_{n})=-\alpha _{0},  \notag
\end{gather}%
$\delta _{il}$ being the Kronecker's symbol.

To further simplify Eq. (\ref{v16}), let us group the terms into pairs, so
that the arguments of the corresponding exponents differ only by sign. For
example, the first pair is formed of the terms with coefficients%
\begin{equation*}
a_{0}=1
\end{equation*}%
and%
\begin{gather*}
a_{123...N}\equiv A_{0}^{2}= \\
\left\vert 
\begin{array}{cccc}
1 & \frac{2\sqrt{\kappa _{1}\kappa _{2}}}{\kappa _{1}+\kappa _{2}} & ... & 
\frac{2\sqrt{\kappa _{1}\kappa _{N}}}{\kappa _{1}+\kappa _{N}} \\ 
\frac{2\sqrt{\kappa _{1}\kappa _{2}}}{\kappa _{1}+\kappa _{2}} & 1 & ... & 
\frac{2\sqrt{\kappa _{2}\kappa _{N}}}{\kappa _{2}+\kappa _{N}} \\ 
... & ... & ... & ... \\ 
\frac{2\sqrt{\kappa _{1}\kappa _{N}}}{\kappa _{1}+\kappa _{N}} & \frac{2%
\sqrt{\kappa _{2}\kappa _{N}}}{\kappa _{2}+\kappa _{N}} & ... & 1%
\end{array}%
\right\vert .
\end{gather*}%
Note that%
\begin{equation*}
\det \left( \tilde{B}_{N}\right) =a_{0}\exp (\alpha _{0}),\ \det \left( 
\tilde{C}_{N}\right) =a_{123...N}\exp (-\alpha _{0}).
\end{equation*}%
Here we defined a new coefficient $A_{0}$, whose subscript "0"\ emphasizes
that the expression $-\alpha _{0}=-\kappa _{1}(x-x_{1})-\kappa
_{2}(x-x_{2})-...-\kappa _{N}(x-x_{N})$ contains no terms (0 terms) with
plus sign. The same logic can be applied to all terms of Eq. (\ref{v16}).
For example, the appropriate partner for the term $a_{1}\exp (\alpha _{1})$
is $a_{234...N}\exp (-\alpha _{1}),$ where%
\begin{equation*}
a_{1}=1,
\end{equation*}%
\begin{gather*}
a_{234...N}\equiv A_{1}^{2}= \\
\left\vert 
\begin{array}{cccc}
1 & \frac{2\sqrt{\kappa _{2}\kappa _{3}}}{\kappa _{2}+\kappa _{3}} & ... & 
\frac{2\sqrt{\kappa _{2}\kappa _{N}}}{\kappa _{2}+\kappa _{N}} \\ 
\frac{2\sqrt{\kappa _{2}\kappa _{3}}}{\kappa _{2}+\kappa _{3}} & 1 & ... & 
\frac{2\sqrt{\kappa _{3}\kappa _{N}}}{\kappa _{3}+\kappa _{N}} \\ 
... & ... & ... & ... \\ 
\frac{2\sqrt{\kappa _{2}\kappa _{N}}}{\kappa _{2}+\kappa _{N}} & \frac{2%
\sqrt{\kappa _{3}\kappa _{N}}}{\kappa _{3}+\kappa _{N}} & ... & 1%
\end{array}%
\right\vert .
\end{gather*}%
Analogously, we can form a pair from $a_{12}\exp (\alpha _{12})$ and $%
a_{345...N}\exp (-\alpha _{12})$, where 
\begin{equation*}
a_{12}=\left\vert 
\begin{array}{cc}
1 & \frac{2\sqrt{\kappa _{1}\kappa _{2}}}{\kappa _{1}+\kappa _{2}} \\ 
\frac{2\sqrt{\kappa _{1}\kappa _{2}}}{\kappa _{1}+\kappa _{2}} & 1%
\end{array}%
\right\vert ,
\end{equation*}%
\begin{equation*}
a_{345...N}=\left\vert 
\begin{array}{cccc}
1 & \frac{2\sqrt{\kappa _{3}\kappa _{4}}}{\kappa _{3}+\kappa _{4}} & ... & 
\frac{2\sqrt{\kappa _{3}\kappa _{N}}}{\kappa _{3}+\kappa _{N}} \\ 
\frac{2\sqrt{\kappa _{3}\kappa _{4}}}{\kappa _{3}+\kappa _{4}} & 1 & ... & 
\frac{2\sqrt{\kappa _{4}\kappa _{N}}}{\kappa _{4}+\kappa _{N}} \\ 
... & ... & ... & ... \\ 
\frac{2\sqrt{\kappa _{3}\kappa _{N}}}{\kappa _{3}+\kappa _{N}} & \frac{2%
\sqrt{\kappa _{4}\kappa _{N}}}{\kappa _{4}+\kappa _{N}} & ... & 1%
\end{array}%
\right\vert .
\end{equation*}%
The principle is simple: for any $a_{n_{1}n_{2}...n_{i}}\exp (\alpha
_{n_{1}n_{2}...n_{i}})$ the partner is $a_{n_{i+1}n_{i+2}...n_{N}}\exp
(-\alpha _{n_{1}n_{2}...n_{i}}).$ In addition, as we will see below, it is
convenient to define a relevant coefficient%
\begin{equation}
A_{n_{1}n_{2}...n_{i}}^{2}\equiv a_{n_{1}n_{2}...n_{i}}\cdot
a_{n_{i+1}n_{i+2}...n_{N}},  \label{v18}
\end{equation}%
where the indices point at the terms with plus sign on the right side of the
expression%
\begin{gather}
-\alpha _{n_{1}...n_{i}}=\kappa _{n_{1}}(x-x_{n_{1}})+...  \label{v19} \\
+\kappa _{n_{i}}(x-x_{n_{i}})-\kappa _{n_{i+1}}(x-x_{n_{i+1}})-...  \notag
\end{gather}%
Looking at the structure of the matrix $\tilde{A}_{N}$, it is obvious that
all these plus sign terms can only originate from the expansion of $\det
\left( \tilde{B}_{N}\right) $ and they correspond to the product $(\tilde{B}%
_{N})_{n_{1}n_{1}}\cdot (\tilde{B}_{N})_{n_{2}n_{2}}\cdot ...\cdot (\tilde{B}%
_{N})_{n_{i}n_{i}}.$ The terms with minus sign on the right side of Eq. (\ref%
{v19}) are related to the expansion of $\det \left( \tilde{C}_{N}\right) ,$
without any contribution from $\det \left( \tilde{B}_{N}\right) .$

On the basis of the above arguments, the following conclusions can be made:

\begin{itemize}
\item All terms on the right side of Eq. (\ref{v16}) can be grouped into
pairs. There is only one term, $a_{123...N}\exp (-\alpha _{0})=\det \left( 
\tilde{C}_{N}\right) $ (with partner $\det \left( \tilde{B}_{N}\right)
=a_{0}\exp (\alpha _{0})$) which is entirely formed of the elements of the
matrix $\tilde{C}_{N}$. Any other term (both partners) contains some
diagonal elements of the matrix $\tilde{B}_{N}$ as well.

\item Any term $a_{n_{1}n_{2}...n_{i}}\exp (\alpha _{n_{1}n_{2}...n_{i}})$ ($%
i=1,2,...$) can be obtained by replacing all elements of the rows and
columns $n_{i+1},n_{i+2},...n_{N}$ of the matrix $\tilde{C}_{N}$ with
corresponding elements of the matrix $\tilde{B}_{N}$ (mostly with zeros). As
a result, one gets a modified matrix $\widehat{C}_{N}$, while $%
a_{n_{1}n_{2}...n_{i}}\exp (\alpha _{n_{1}n_{2}...n_{i}})=\det \left( 
\widehat{C}_{N}\right) .$ The Laplace expansion (\ref{v15}) of this
determinant for the fixed rows $n_{i+1},n_{i+2},...n_{N}$ contains only one
term!

\item For any term $a_{n_{1}n_{2}...n_{i}}\exp (\alpha
_{n_{1}n_{2}...n_{i}}) $ of the expansion (\ref{v16}) there is a partner%
\begin{equation*}
a_{n_{i+1}n_{i+2}...n_{N}}\exp (-\alpha _{n_{1}n_{2}...n_{i}})=\det \left( 
\overline{C}_{N}\right) ,
\end{equation*}%
where $\overline{C}_{N}$ can be obtained by replacing all elements of the
rows and columns $n_{1},n_{2},...n_{i}$ of the matrix $\tilde{C}_{N}$ with
corresponding elements of the matrix $\tilde{B}_{N}$. The Laplace expansion
of $\det \left( \overline{C}_{N}\right) $ for the rows $n_{1},n_{2},...n_{i}$
also contains only one term: 
\begin{gather}
\det \left( \overline{C}_{N}\right) =\exp (-\alpha _{12...i})\times
\label{v20} \\
\left\vert 
\begin{array}{cccc}
1 & \frac{2\sqrt{\kappa _{n_{i+1}}\kappa _{n_{i+2}}}}{\kappa
_{n_{i+1}}+\kappa _{n_{i+2}}} & ... & \frac{2\sqrt{\kappa _{n_{i+1}}\kappa
_{n_{N}}}}{\kappa _{n_{i+1}+}\kappa _{n_{N}}} \\ 
\frac{2\sqrt{\kappa _{n_{i+1}}\kappa _{n_{i+2}}}}{\kappa _{n_{i+1}}+\kappa
_{n_{i+2}}} & 1 & ... & \frac{2\sqrt{\kappa _{n_{i+2}}\kappa _{n_{N}}}}{%
\kappa _{n_{i+2}+}\kappa _{n_{N}}} \\ 
... & ... & ... & ... \\ 
\frac{2\sqrt{\kappa _{n_{i+1}}\kappa _{n_{N}}}}{\kappa _{n_{i+1}+}\kappa
_{n_{N}}} & \frac{2\sqrt{\kappa _{n_{i+2}}\kappa _{n_{N}}}}{\kappa
_{n_{i+2}+}\kappa _{n_{N}}} & ... & 1%
\end{array}%
\right\vert .  \notag
\end{gather}%
It means, for example, that%
\begin{gather}
A_{12...i}^{2}=a_{12...i}\cdot a_{i+1,i+2,...N}=  \label{v21} \\
\left\vert 
\begin{array}{cccc}
1 & \frac{2\sqrt{\kappa _{1}\kappa _{2}}}{\kappa _{1}+\kappa _{2}} & ... & 
\frac{2\sqrt{\kappa _{1}\kappa _{i}}}{\kappa _{1}+\kappa _{i}} \\ 
\frac{2\sqrt{\kappa _{1}\kappa _{2}}}{\kappa _{1}+\kappa _{2}} & 1 & ... & 
\frac{2\sqrt{\kappa _{2}\kappa _{i}}}{\kappa _{2}+\kappa _{i}} \\ 
... & ... & ... & ... \\ 
\frac{2\sqrt{\kappa _{1}\kappa _{i}}}{\kappa _{1}+\kappa _{i}} & \frac{2%
\sqrt{\kappa _{2}\kappa _{i}}}{\kappa _{2}+\kappa _{i}} & ... & 1%
\end{array}%
\right\vert \times  \notag \\
\left\vert 
\begin{array}{cccc}
1 & \frac{2\sqrt{\kappa _{i+1}\kappa _{i+2}}}{\kappa _{i+1}+\kappa _{i+2}} & 
... & \frac{2\sqrt{\kappa _{i+1}\kappa _{N}}}{\kappa _{i+1}+\kappa _{N}} \\ 
\frac{2\sqrt{\kappa _{i+1}\kappa _{i+2}}}{\kappa _{i+1}+\kappa _{i+2}} & 1 & 
... & \frac{2\sqrt{\kappa _{n_{i+2}}\kappa _{n_{N}}}}{\kappa
_{n_{i+2}+}\kappa _{n_{N}}} \\ 
... & ... & ... & ... \\ 
\frac{2\sqrt{\kappa _{i+1}\kappa _{N}}}{\kappa _{i+1}+\kappa _{N}} & \frac{2%
\sqrt{\kappa _{n_{i+2}}\kappa _{n_{N}}}}{\kappa _{n_{i+2}+}\kappa _{n_{N}}}
& ... & 1%
\end{array}%
\right\vert .  \notag
\end{gather}

\item One should avoid re-use of the terms: an already existing pair must
not be included again! It means that the members of the modified expansion (%
\ref{v16}) are identified by no more than $\left[ N/2\right] $ indices
(square brackets denote integer part of $N/2$). It is convenient to group
the members on the basis of the number of indices, so that there will be $%
0,1,2,...,$ $\left[ N/2\right] $ different indices.
\end{itemize}

As the final result of the above analysis, we get the following general
formula:%
\begin{gather}
\tau _{N}=\det \left( \tilde{A}_{N}\right) =\sum\limits_{i=0}^{N}A_{i}\left[
\exp (\alpha _{i}+\beta _{i})+\exp (-\alpha _{i}-\beta _{i})\right] +  \notag
\\
...+\sum\limits_{1\leq i_{1}<...<i_{\left[ N/2\right] }\leq N}A_{i_{1}...i_{%
\left[ N/2\right] }}[\exp (\alpha _{i_{1}...i_{\left[ N/2\right] }}+\beta
_{i_{1}...i_{\left[ N/2\right] }})+  \notag \\
\exp (-\alpha _{i_{1}..i_{\left[ N/2\right] }}-\beta _{i_{1}...i_{\left[ N/2%
\right] }})]=\sum\limits_{i=0}^{N}2A_{i}\cosh (\alpha _{i}+\beta _{i})+ 
\notag \\
\sum\limits_{1\leq i_{1}<i_{2}\leq N}2A_{i_{1}i_{2}}\cosh (\alpha
_{i_{1}i_{2}}+\beta _{i_{1}i_{2}})+...+  \label{v22} \\
\sum\limits_{1\leq i_{1}<...<i_{\left[ N/2\right] }\leq
N}2A_{i_{1}i_{2}...i_{\left[ N/2\right] }}\cosh (\alpha _{i_{1}i_{2}...i_{%
\left[ N/2\right] }}+\beta _{i_{1}i_{2}...i_{\left[ N/2\right] }}),  \notag
\end{gather}%
where%
\begin{equation}
\exp (\beta _{n_{1}n_{2}...n_{i}})\equiv \dfrac{a_{n_{1}n_{2}...n_{i}}}{%
a_{n_{i+1}n_{i+2}...n_{N}}}.  \label{v23}
\end{equation}

It is easy to be convinced that the expansion (\ref{v22}) contains $2^{N-1}$
members in total (apart from inessential factor $2$). Indeed, according to
Newton's binomial theorem%
\begin{equation*}
\begin{pmatrix}
N \\ 
0%
\end{pmatrix}%
+%
\begin{pmatrix}
N \\ 
1%
\end{pmatrix}%
+%
\begin{pmatrix}
N \\ 
2%
\end{pmatrix}%
+%
\begin{pmatrix}
N \\ 
3%
\end{pmatrix}%
+...+=\frac{\left( 1+1\right) ^{N}}{2}=2^{N-1},
\end{equation*}%
where we took into consideration that only half of this formal series is
actually needed.

\section{Alternants of $\protect\tau $-functions}

We have shown that not only $\det \left( \tilde{A}_{N}\right) $ itself but
also the coefficients $A_{i},A_{i_{1}i_{2}},...,A_{i_{1}i_{2}...i_{\left[ N/2%
\right] }}$ in the expansion (\ref{v22}) are $\tau $-functions.
Consequently, the solution of the inverse scattering problem has been
reduced to evaluating a number of determinants 
\begin{equation}
D(\kappa _{1},\kappa _{2},..,\kappa _{n})\equiv \left\vert 
\begin{array}{cccc}
1 & \frac{2\sqrt{\kappa _{1}\kappa _{2}}}{\kappa _{1}+\kappa _{2}} & ... & 
\frac{2\sqrt{\kappa _{1}\kappa _{n}}}{\kappa _{1}+\kappa _{n}} \\ 
\frac{2\sqrt{\kappa _{1}\kappa _{2}}}{\kappa _{1}+\kappa _{2}} & 1 & ... & 
\frac{2\sqrt{\kappa _{2}\kappa _{n}}}{\kappa _{2}+\kappa _{n}} \\ 
... & ... & ... & ... \\ 
\frac{2\sqrt{\kappa _{1}\kappa _{n}}}{\kappa _{1}+\kappa _{n}} & \frac{2%
\sqrt{\kappa _{2}\kappa _{n}}}{\kappa _{2}+\kappa _{n}} & ... & 1%
\end{array}%
\right\vert ,  \label{Alt1}
\end{equation}%
fixed by the parameters $\kappa _{1}<\kappa _{2}<$ $...<\kappa _{n}$, $n$
being an appropriate natural number. We are now going to derive a simple
formula for calculating such $\tau $-functions. First, let us set a
one-to-one correspondence between each row of the determinant and a fixed
parameter%
\begin{equation}
q_{i}=\sqrt{\kappa _{i}}\ \left( i=1,2,...,n\right) .  \label{Alt2}
\end{equation}%
For example, the modified elements of the first row of Eq. (\ref{Alt1}) will
be%
\begin{equation*}
1=\frac{2q_{1}\sqrt{\kappa _{1}}}{q_{1}^{2}+\kappa _{1}},\text{ }\frac{2%
\sqrt{\kappa _{1}\kappa _{2}}}{\kappa _{1}+\kappa _{2}}=\frac{2q_{1}\sqrt{%
\kappa _{2}}}{q_{1}^{2}+\kappa _{2}},...,\frac{2\sqrt{\kappa _{1}\kappa _{n}}%
}{\kappa _{1}+\kappa _{n}}=\frac{2q_{1}\sqrt{\kappa _{n}}}{q_{1}^{2}+\kappa
_{n}}.
\end{equation*}%
The usefulness of this trick soon becomes evident, although there seems to
be only a formal change:%
\begin{gather}
D(\kappa _{1},\kappa _{2},..,\kappa _{n})\equiv \left\vert 
\begin{array}{cccc}
\frac{2q_{1}\sqrt{\kappa _{1}}}{q_{1}^{2}+\kappa _{1}} & \frac{2q_{1}\sqrt{%
\kappa _{2}}}{q_{1}^{2}+\kappa _{2}} & ... & \frac{2q_{1}\sqrt{\kappa _{n}}}{%
q_{1}^{2}+\kappa _{n}} \\ 
\frac{2q_{2}\sqrt{\kappa _{1}}}{q_{2}^{2}+\kappa _{1}} & \frac{2q_{2}\sqrt{%
\kappa _{2}}}{q_{2}^{2}+\kappa _{2}} & ... & \frac{2q_{2}\sqrt{\kappa _{n}}}{%
q_{2}^{2}+\kappa _{n}} \\ 
... & ... & ... & ... \\ 
\frac{2q_{n}\sqrt{\kappa _{1}}}{q_{n}^{2}+\kappa _{1}} & \frac{2q_{n}\sqrt{%
\kappa _{2}}}{q_{n}^{2}+\kappa _{2}} & ... & \frac{2q_{n}\sqrt{\kappa _{n}}}{%
q_{n}^{2}+\kappa _{n}}%
\end{array}%
\right\vert =  \notag \\
2^{n}\cdot \kappa _{1}\kappa _{2}...\kappa _{n}\left\vert 
\begin{array}{cccc}
\frac{1}{q_{1}^{2}+\kappa _{1}} & \frac{1}{q_{1}^{2}+\kappa _{2}} & ... & 
\frac{1}{q_{1}^{2}+\kappa _{n}} \\ 
\frac{1}{q_{2}^{2}+\kappa _{1}} & \frac{1}{q_{2}^{2}+\kappa _{2}} & ... & 
\frac{1}{q_{2}^{2}+\kappa _{n}} \\ 
... & ... & ... & ... \\ 
\frac{1}{q_{n}^{2}+\kappa _{1}} & \frac{1}{q_{n}^{2}+\kappa _{2}} & ... & 
\frac{1}{q_{n}^{2}+\kappa _{n}}%
\end{array}%
\right\vert .  \label{Alt3}
\end{gather}%
Here we separated a factor $2q_{i}$ from each row and $\sqrt{\kappa _{i}}$
from each column, i.e., $2q_{i}\sqrt{\kappa _{i}}=2\kappa _{i}$ ($%
i=1,2,...,n $) from each such pair.

Next, let us transform Eq. (\ref{Alt3}) into a polynomial, multiplying each
row by%
\begin{equation*}
\prod\limits_{j}\left( q_{i}^{2}+\kappa _{j}\right) \ \left(
i=1,2,...,n\right) .
\end{equation*}%
The result is%
\begin{equation}
D(\kappa _{1},\kappa _{2},..,\kappa _{n})=\dfrac{1}{\prod\limits_{1\leq
i<j\leq n}\left( \kappa _{i}+\kappa _{j}\right) ^{2}}\times D_{n},
\label{Alt4}
\end{equation}%
where a new determinant%
\begin{equation}
D_{n}\equiv \left\vert 
\begin{array}{cccc}
\prod\limits_{j\neq 1}\left( q_{1}^{2}+\kappa _{j}\right) & 
\prod\limits_{j\neq 2}\left( q_{1}^{2}+\kappa _{j}\right) & ... & 
\prod\limits_{j\neq n}\left( q_{1}^{2}+\kappa _{j}\right) \\ 
\prod\limits_{j\neq 1}\left( q_{2}^{2}+\kappa _{j}\right) & 
\prod\limits_{j\neq 2}\left( q_{2}^{2}+\kappa _{j}\right) & ... & 
\prod\limits_{j\neq n}\left( q_{2}^{2}+\kappa _{j}\right) \\ 
... & ... & ... & ... \\ 
\prod\limits_{j\neq 1}\left( q_{n}^{2}+\kappa _{j}\right) & 
\prod\limits_{j\neq 2}\left( q_{n}^{2}+\kappa _{j}\right) & ... & 
\prod\limits_{j\neq n}\left( q_{n}^{2}+\kappa _{j}\right)%
\end{array}%
\right\vert  \label{Alt5}
\end{equation}%
was introduced. As can be seen, the factor $2^{n}\cdot \kappa _{1}\kappa
_{2}...\kappa _{n}$ was canceled out from Eq. (\ref{Alt4}).

We can see that the elements of the columns of $D_{n}$ correspond to
different values of the same function, while any row is characterized by a
single fixed parameter. Indeed, Eq. (\ref{Alt5}) can be expressed as%
\begin{equation}
D_{n}=\left\vert 
\begin{array}{cccc}
F_{1}(q_{1}) & F_{2}(q_{1}) & ... & F_{n}(q_{1}) \\ 
F_{1}(q_{2}) & F_{2}(q_{2}) & ... & F_{n}(q_{2}) \\ 
... & ... & ... & ... \\ 
F_{1}(q_{n}) & F_{2}(q_{n}) & ... & F_{n}(q_{n})%
\end{array}%
\right\vert,  \label{Alt6}
\end{equation}%
where%
\begin{eqnarray}
F_{1}(x) &\equiv &\left( x^{2}+\kappa _{2}\right) \left( x^{2}+\kappa
_{3}\right) ...\left( x^{2}+\kappa _{n}\right),  \label{Alt7} \\
F_{2}(x) &\equiv &\left( x^{2}+\kappa _{1}\right) \left( x^{2}+\kappa
_{3}\right) ...\left( x^{2}+\kappa _{n}\right),  \notag \\
F_{3}(x) &\equiv &\left( x^{2}+\kappa _{1}\right) \left( x^{2}+\kappa
_{2}\right) \left( x^{2}+\kappa _{4}\right) ...\left( x^{2}+\kappa
_{n}\right),  \notag \\
&&...  \notag \\
F_{n}(x) &\equiv &\left( x^{2}+\kappa _{1}\right) \left( x^{2}+\kappa
_{2}\right) \left( x^{2}+\kappa _{3}\right) ...\left( x^{2}+\kappa
_{n-1}\right).  \notag
\end{eqnarray}
A determinant that has a structure of Eq. (\ref{Alt6}) is called \textit{%
alternant} (see \cite{Muir}, p. 161). The best known alternant is
Vandermonde's determinant (for the same set of variables) 
\begin{equation}
V_{n}\equiv \left\vert 
\begin{array}{ccccc}
1 & q_{1} & q_{1}^{2} & ... & q_{1}^{n-1} \\ 
1 & q_{2} & q_{2}^{2} & ... & q_{2}^{n-1} \\ 
... & ... & ... & ... & ... \\ 
1 & q_{n-1} & q_{n-1}^{2} & ... & q_{n-1}^{n-1} \\ 
1 & q_{n} & q_{n}^{2} & ... & q_{n}^{n-1}%
\end{array}%
\right\vert ,  \label{Alt8}
\end{equation}%
which can be easily evaluated (see \cite{Korn}, p. 16):
\begin{equation}
V_{n}(q_{1},q_{2},...,q_{n})=\prod\limits_{1\leq i<j\leq n}\left(
q_{j}-q_{i}\right) .  \label{Alt9}
\end{equation}
An important point is that the factor $V_{n}(q_{1},q_{2},...,q_{n})$ can be
separated from any $n$th-order alternant. Indeed, the argument $q_{n}$ may
only appear in the $n$th row of Eq. (\ref{Alt6}): if we put it into any
other row then the determinant would be identically zero. It means that $%
D_{n}$ has a factor $\prod\limits_{i=1}^{n-1}\left( q_{n}-q_{i}\right) .$
Analogous reasoning applied to $q_{n-1}$ shows that $D_{n}$ also has a
factor $\prod\limits_{i=1}^{n-2}\left( q_{n-1}-q_{i}\right) $, etc. Putting
it all together, we conclude that an $n$th-order alternate always has a
factor $V_{n}(q_{1},q_{2},...,q_{n}).$

To continue the analysis, let us recall some useful properties of the
elementary symmetric functions:%
\begin{eqnarray}
\sigma _{0} &\equiv &1,  \label{Alt10} \\
\sigma _{1} &\equiv &q_{1}+q_{2}+...+q_{n},  \notag \\
&&...  \notag \\
\sigma _{k} &\equiv &\sum_{1\leq i_{1}<i_{2}\cdot \cdot \cdot <i_{k}\leq
n}q_{i_{1}}q_{i_{2}}\cdot \cdot \cdot \ q_{i_{k}},  \notag \\
&&...  \notag \\
\sigma _{n} &\equiv &q_{1}q_{2}\cdot \cdot \cdot \ q_{n}.  \notag
\end{eqnarray}%
Here $\sigma _{k}$ ($k\geq 2$) is a sum of all possible products of exactly $%
k$ variables arranged in ascending order of their indices. According to the
Fundamental Theorem for symmetric polynomials (see \cite{CLO}, p. 312), any
such polynomial can be uniquely expressed as a polynomial in $\sigma
_{1},\sigma _{2},...,\sigma _{n}.$ This in turn is a basis for the following
important theorem:

\textbf{Theorem 1}: Let $\left\vert A_{n}\right\vert $ be an $n$th order
alternant generated by the functions%
\begin{eqnarray}
F_{j}(x) &=&a_{0j}+a_{1j}\cdot x+a_{2j}\cdot x^{2}+...+a_{nj}\cdot x^{n},
\label{Alt106} \\
\ j &=&1,2,...,n,  \notag
\end{eqnarray}
where the parameters $a_{ij}$ do not depend on $x$, and define%
\begin{equation}
S_{k}\equiv \left( -1\right) ^{k}\sigma _{k}(q_{1},q_{2},...,q_{n}).
\label{Alt107}
\end{equation}%
Then%
\begin{equation}
\dfrac{\left\vert A_{n}\right\vert }{V_{n}(q_{1},q_{2},...,q_{n})}%
=\left\vert 
\begin{array}{ccccc}
a_{01} & a_{11} & a_{21} & ... & a_{n1} \\ 
a_{02} & a_{12} & a_{22} & ... & a_{n2} \\ 
... & ... & ... & ... & ... \\ 
a_{0n} & a_{1n} & a_{2n} & ... & a_{nn} \\ 
S_{n} & S_{n-1} & S_{n-2} & ... & S_{0}%
\end{array}%
\right\vert \equiv \Delta _{n+1}.  \label{Alt11}
\end{equation}

\textbf{Proof:} Let us introduce an auxiliary polynomial 
\begin{equation}
F(x)\equiv (x-q_{1})(x-q_{2})...(x-q_{n})  \label{Alt105}
\end{equation}%
and form an $(n+1)$th order Vandermonde's determinant, adding a new
(arbitrary) variable $q_{n+1}$, so that%
\begin{equation*}
V_{n+1}(q_{1},q_{2},...,q_{n},q_{n+1})=\left\vert 
\begin{array}{ccccc}
1 & 1 & ... & 1 & 1 \\ 
q_{1} & q_{2} & ... & q_{n} & q_{n+1} \\ 
... & ... & ... & ... & ... \\ 
q_{1}^{n-1} & q_{2}^{n-1} & ... & q_{n}^{n-1} & q_{n+1}^{n-1} \\ 
q_{1}^{n} & q_{2}^{n} & ... & q_{n}^{n} & q_{n+1}^{n}%
\end{array}%
\right\vert
\end{equation*}%
(compared with Eq. (\ref{Alt8}), the row and column indices are
interchanged). Multiplying $\Delta _{n+1}\ $by $V_{n+1}$ and using Eqs. (\ref%
{Alt106}), (\ref{Alt107}), (\ref{Alt105}), we get 
\begin{equation*}
\Delta _{n+1}\times V_{n+1}=\left\vert 
\begin{array}{ccccc}
F_{1}(q_{1}) & F_{1}(q_{2}) & ... & F_{1}(q_{n}) & F_{1}(q_{n+1}) \\ 
F_{2}(q_{1}) & F_{2}(q_{2}) & ... & F_{2}(q_{n}) & F_{2}(q_{n+1}) \\ 
... & ... & ... & ... & ... \\ 
F_{n}(q_{1}) & F_{n}(q_{2}) & ... & F_{n}(q_{n}) & F_{n}(q_{n+1}) \\ 
F(q_{1}) & F(q_{2}) & ... & F(q_{n}) & F(q_{n+1})%
\end{array}%
\right\vert =
\end{equation*}%
\begin{equation*}
\left\vert 
\begin{array}{ccccc}
F_{1}(q_{1}) & F_{1}(q_{2}) & ... & F_{1}(q_{n}) & F_{1}(q_{n+1}) \\ 
F_{2}(q_{1}) & F_{2}(q_{2}) & ... & F_{2}(q_{n}) & F_{2}(q_{n+1}) \\ 
... & ... & ... & ... & ... \\ 
F_{n}(q_{1}) & F_{n}(q_{2}) & ... & F_{n}(q_{n}) & F_{n}(q_{n+1}) \\ 
0 & 0 & ... & 0 & F(q_{n+1})%
\end{array}%
\right\vert
\end{equation*}%
or, in a more compact form, 
\begin{equation}
\Delta _{n+1}\times V_{n+1}=(q_{n+1}-q_{1})...(q_{n+1}-q_{n})\times
\left\vert A_{n}\right\vert ,  \label{Alt12}
\end{equation}%
where we took into consideration that $F(q_{i})=0$ if $i=1,2,...,n.$

From Eq. (\ref{Alt9}) one concludes that%
\begin{equation*}
V_{n+1}=(q_{n+1}-q_{1})(q_{n+1}-q_{2})...(q_{n+1}-q_{n})\cdot V_{n},
\end{equation*}%
so that both sides of Eq. (\ref{Alt12}) contain a common divisor $%
(q_{n+1}-q_{1})(q_{n+1}-q_{2})...(q_{n+1}-q_{n})\neq 0$. Consequently, $%
\Delta _{n+1}V_{n}=\left\vert A_{n}\right\vert ,$ Q. E. D.

\subsection{Relationship to the inverse scattering problem}

Let us apply Theorem 1 to the alternant (\ref{Alt6}) generated by the
functions (\ref{Alt7}), which are polynomials in a variable $x^{2}$. Using
Eqs. (\ref{Alt106}) and (\ref{Alt107}), one gets%
\begin{eqnarray*}
a_{01} &=&\sigma _{n-1}(\kappa _{2},\kappa _{3},...,\kappa _{n}), \\
\ a_{11} &=&\sigma _{n-2}(\kappa _{2},\kappa _{3},...,\kappa
_{n}),...,a_{n-1,1}=1,\ a_{n1}=0, \\
a_{02} &=&\sigma _{n-1}(\kappa _{1},\kappa _{3},...,\kappa _{n}), \\
\ a_{12} &=&\sigma _{n-2}(\kappa _{1},\kappa _{3},...,\kappa
_{n}),...,a_{n-1,2}=1,\ a_{n2}=0, \\
&&... \\
a_{0n} &=&\sigma _{n-1}(\kappa _{1},\kappa _{2},...,\kappa _{n-1}), \\
a_{1n} &=&\sigma _{n-2}(\kappa _{1},\kappa _{2},...,\kappa
_{n-1}),...,a_{n-1,n}=1,\ a_{nn}=0.
\end{eqnarray*}%
Consequently,%
\begin{gather}
D_{n}=V_{n}(q_{1},q_{2},...,q_{n})\times \left\vert R_{n}\right\vert
,~\left\vert R_{n}\right\vert \equiv  \label{Alt13} \\
\left\vert 
\begin{array}{ccccc}
\sigma _{n-1}(\kappa _{2},\kappa _{3},...,\kappa _{n}) & \sigma
_{n-2}(\kappa _{2},\kappa _{3},...,\kappa _{n}) & ... & 1 & 0 \\ 
\sigma _{n-1}(\kappa _{1},\kappa _{3},...,\kappa _{n}) & \sigma
_{n-2}(\kappa _{1},\kappa _{3},...,\kappa _{n}) & ... & 1 & 0 \\ 
... & ... & ... & ... & ... \\ 
\sigma _{n-1}(\kappa _{1},\kappa _{2},...,\kappa _{n-1}) & \sigma
_{n-2}(\kappa _{1},\kappa _{2},...,\kappa _{n-1}) & ... & 1 & 0 \\ 
S_{n} & S_{n-1} & ... & S_{1} & S_{0}%
\end{array}%
\right\vert =  \notag \\
\left\vert 
\begin{array}{cccc}
\sigma _{n-1}(\kappa _{2},\kappa _{3},...,\kappa _{n}) & \sigma
_{n-2}(\kappa _{2},\kappa _{3},...,\kappa _{n}) & ... & 1 \\ 
\sigma _{n-1}(\kappa _{1},\kappa _{3},...,\kappa _{n}) & \sigma
_{n-2}(\kappa _{1},\kappa _{3},...,\kappa _{n}) & ... & 1 \\ 
... & ... & ... & ... \\ 
\sigma _{n-1}(\kappa _{1},\kappa _{2},...,\kappa _{n-1}) & \sigma
_{n-2}(\kappa _{1},\kappa _{2},...,\kappa _{n-1}) & ... & 1%
\end{array}%
\right\vert .  \notag
\end{gather}

Here we defined a new determinant $\left\vert R_{n}\right\vert $ which seems
to be another alternant, so we can apply Theorem 1 to evaluate it. To be
convinced that $\left\vert R_{n}\right\vert $ is indeed an alternant, we
have to specify the generating functions. Obviously,%
\begin{eqnarray}
F_{n}(x) &=&1=\sigma _{0}(\kappa _{1},\kappa _{2},...,\kappa _{n}),
\label{Alt14} \\
F_{n-1}(x) &=&\sigma _{1}(\kappa _{1},\kappa _{2},...,\kappa _{n})-x.  \notag
\end{eqnarray}%
Also, it is easy to prove that 
\begin{equation}
F_{n-2}(x)=\sigma _{2}-xF_{n-1}(x)=\sigma _{2}-\sigma _{1}x+x^{2},
\label{Alt15}
\end{equation}%
where we dropped the arguments to get a more compact formula. Let us agree
that if no arguments are explicitly given then the corresponding function
depends on $n$ arguments: $\kappa _{1},\kappa _{2},...,\kappa _{n}$.

Taking, e.g., $x=\kappa _{1}$ we can check the validity of Eq. (\ref{Alt15}%
). Indeed,%
\begin{gather*}
F_{n-2}(\kappa _{1})=\sigma _{2}-\sigma _{1}\kappa _{1}+\kappa _{1}^{2}= \\
\sum_{1\leq i_{1}<i_{2}\leq n}\left( \kappa _{i_{1}}\kappa _{i_{2}}\right)
-\kappa _{1}\left( \kappa _{1}+\kappa _{2}+...+\kappa _{n}\right) +\kappa
_{1}^{2}= \\
\sigma _{n-1}(\kappa _{2},\kappa _{3},...,\kappa _{n}),
\end{gather*}%
as needed according to Eq. (\ref{Alt13}). Continuing in the same manner, we
get the following result:%
\begin{eqnarray}
F_{n-3}(x) &=&\sigma _{3}-xF_{n-2}(x)=\sigma _{3}-\sigma _{2}x+\sigma
_{1}x^{2}-x^{3},  \notag \\
&&...  \notag \\
F_{n-i}(x) &=&\sigma _{n-i}-xF_{n-i-1}(x)=\sum_{j=0}^{i}(-1)^{j}\sigma
_{i-j}x^{j}\rightarrow  \notag \\
F_{k}(x) &=&\sum_{j=0}^{n-k}(-1)^{j}\sigma
_{n-k-j}x^{j}=\sum_{j=0}^{n-k}a_{jk}x^{j},  \label{Alt16} \\
a_{jk} &=&\left\{ 
\begin{array}{l}
(-1)^{j}\sigma _{n-k-j},\text{ if }j\leq n-k, \\ 
0,\text{ if }j>n-k.%
\end{array}%
\right.  \label{Alt17}
\end{eqnarray}%
Thus $\left\vert R_{n}\right\vert $ is indeed an alternant with generating
functions (\ref{Alt106}). Consequently, according to Eqs. (\ref{Alt107}), (%
\ref{Alt11}) and (\ref{Alt17}),%
\begin{equation*}
\frac{\left\vert R_{n}\right\vert }{V_{n}}=\left\vert 
\begin{array}{ccccc}
a_{01} & a_{11} & a_{21} & ... & a_{n1} \\ 
a_{02} & a_{12} & a_{22} & ... & a_{n2} \\ 
... & ... & ... & ... & ... \\ 
a_{0n} & a_{1n} & a_{2n} & ... & a_{nn} \\ 
S_{n} & S_{n-1} & S_{n-2} & ... & S_{0}%
\end{array}%
\right\vert =
\end{equation*}%
%
%
%
%
%
%
\begin{equation*}
\left\vert 
\begin{array}{ccccccc}
\sigma _{n-1} & -\sigma _{n-2} & \sigma _{n-3} & ... & \left( -1\right)
^{n-2}\sigma _{1} & \left( -1\right) ^{n-1} & 0 \\ 
\sigma _{n-2} & -\sigma _{n-3} & \sigma _{n-4} & ... & \left( -1\right)
^{n-2} & 0 & 0 \\ 
\sigma _{n-3} & -\sigma _{n-4} & \sigma _{n-5} & ... & 0 & 0 & 0 \\ 
... & ... & ... & ... & 0 & 0 & 0 \\ 
\sigma _{1} & -1 & 0 & ... & 0 & 0 & 0 \\ 
1 & 0 & 0 & ... & 0 & 0 & 0 \\ 
\left( -1\right) ^{n}\sigma _{n} & \left( -1\right) ^{n-1}\sigma _{n-1} & 
\left( -1\right) ^{n-1}\sigma _{n-2} & ... & \sigma _{2} & -\sigma _{1} & 1%
\end{array}%
\right\vert =
\end{equation*}%
%
%
%
%
%
%
\begin{equation}
\left\vert 
\begin{array}{cccccc}
\sigma _{n-1} & -\sigma _{n-2} & \sigma _{n-3} & ... & \left( -1\right)
^{n-2}\sigma _{1} & \left( -1\right) ^{n-1} \\ 
\sigma _{n-2} & -\sigma _{n-3} & \sigma _{n-4} & ... & \left( -1\right)
^{n-2} & 0 \\ 
\sigma _{n-3} & -\sigma _{n-4} & \sigma _{n-5} & ... & 0 & 0 \\ 
... & ... & ... & ... & 0 & 0 \\ 
\sigma _{1} & -1 & 0 & ... & 0 & 0 \\ 
1 & 0 & 0 & ... & 0 & 0%
\end{array}%
\right\vert ,  \label{Alt175}
\end{equation}%
which means that%
\begin{equation}
\dfrac{\left\vert R_{n}\right\vert }{V_{n}}=1.  \label{Alt18}
\end{equation}%
Indeed, we can transform Eq. (\ref{Alt175}), repeatedly using cofactor
expansion in terms of the last column and applying the general definition%
\begin{equation}
\det (A)=\sum\limits_{p}\sigma (p)a_{1p_{1}}a_{2p_{2}}...a_{np_{n}}.
\label{Alt185}
\end{equation}%
Here the sum involves all possible permutations $p=\left(
p_{1},p_{2},...,p_{n}\right) $ of the indices$\ \left( 1,2,...,n\right) ,$ $%
\sigma (p)=(-1)^{N_{p}}$\ and $N_{p}$ is the number of pairwise interchanges
needed to restore the natural order $\left( 1,2,...,n\right)$. For example, $%
\sigma \left( 3,2,1\right) =-1$, but $\sigma \left( 4,3,2,1\right) =1.$ As a
result of the described operation, we obtain%
\begin{equation*}
\dfrac{\left\vert R_{n}\right\vert }{V_{n}}=\sigma (n,n-1,...,1)\cdot (-1)^{%
\left[ n/2\right] }.
\end{equation*}%
On the other hand, $\sigma (n,n-1,...,1)=(-1)^{\left[ n/2\right] }$, since $%
N_{p}=\left[ n/2\right] $. Consequently,%
\begin{equation*}
\left\vert R_{n}\right\vert =V_{n}\cdot (-1)^{\left[ n/2\right] }\cdot (-1)^{%
\left[ n/2\right] }=V_{n}.
\end{equation*}

In summary, we have obtained a very simple and universal recipe for
calculating determinants defined by Eq. (\ref{Alt5}):%
\begin{equation}
D_{n}=\left( V_{n}\right) ^{2}=\prod\limits_{1\leq i<j\leq n}\left( \kappa
_{j}-\kappa _{i}\right) ^{2}.  \label{Alt19}
\end{equation}%
Combining Eqs. (\ref{Alt4}) and (\ref{Alt19}), we can formulate a general
and important result:

\textbf{Theorem 2}: Let $\kappa _{1},\kappa _{2},..,\kappa _{n}$ be
arbitrary positive real numbers arranged in the ascending order, so that $%
\kappa _{1}<\kappa _{2}<$ $...<\kappa _{n},$ and let $D(\kappa _{1},\kappa
_{2},..,\kappa _{n})$ be a determinant, defined by Eq. (\ref{Alt1}). Then%
\begin{equation}
D(\kappa _{1},\kappa _{2},..,\kappa _{n})=\prod\limits_{1\leq i<j\leq
n}\left( \dfrac{\kappa _{j}-\kappa _{i}}{\kappa _{j}+\kappa _{i}}\right)
^{2},  \label{Alt20}
\end{equation}%
where the product contains all possible combinations of the pairs $\left(
\kappa _{j},\kappa _{i}\right) $ satisfying the condition $1\leq i<j\leq n.$

\section{Symmetric reflectionless potentials}

The excursion to theory of determinants concluded with a surprisingly simple
final result. Indeed, as is seen from Eqs. (\ref{v18}), (\ref{v21}) and (\ref%
{v23}), all coefficients in Eq. (\ref{v22}) can be evaluated with the help
of Eq. (\ref{Alt20}), which means that the general formula for $\tau $%
-functions can be essentially simplified. For example,%
\begin{equation}
A_{0}=\sqrt{D(\kappa _{1},\kappa _{2},..,\kappa _{n})}=\prod\limits_{1\leq
i<j\leq n}\left( \frac{\kappa _{j}-\kappa _{i}}{\kappa _{j}+\kappa _{i}}%
\right) ,  \label{Alt21}
\end{equation}%
\begin{equation}
\frac{A_{i}}{A_{0}}=\sqrt{\frac{D(\kappa _{1},\kappa _{2},..,\kappa
_{i-1}\kappa _{i+1}\kappa _{i+2}...\kappa _{n})}{D(\kappa _{1},\kappa
_{2},..,\kappa _{n})}}=\prod\limits_{j\neq i}\left\vert \frac{\kappa
_{j}+\kappa _{i}}{\kappa _{j}-\kappa _{i}}\right\vert ,  \label{Alt22}
\end{equation}%
etc. Thus Eq. (\ref{v22}) transforms to 
\begin{gather}
\tau _{N}=\frac{\det \left( \tilde{A}_{N}\right) }{2A_{0}}=\cosh (\alpha
_{0}+\beta _{0})+\sum_{i}\prod\limits_{j\neq i}\left\vert \frac{\kappa
_{j}+\kappa _{i}}{\kappa _{j}-\kappa _{i}}\right\vert \cosh (\alpha
_{i}+\beta _{i})+  \label{Alt23} \\
\sum_{1\leq i_{1}<i_{2}\leq N}\ \prod\limits_{j\neq i_{1},i_{2}}\left\vert 
\frac{\kappa _{j}+\kappa _{i_{1}}}{\kappa _{j}-\kappa _{i_{1}}}\right\vert
\cdot \left\vert \frac{\kappa _{j}+\kappa _{i_{2}}}{\kappa _{j}-\kappa
_{i_{2}}}\right\vert \cosh (\alpha _{i_{1}i_{2}}+\beta _{i_{1}i_{2}})+...+ 
\notag \\
\sum\limits_{\substack{ 1\leq i_{1}...  \\ <i_{\left[ N/2\right] \leq N}}}\
\prod\limits_{\substack{ j\neq i_{1},i_{2},  \\ ...,i_{\left[ N/2\right] }}}%
\left\vert \dfrac{\kappa _{j}+\kappa _{i_{1}}}{\kappa _{j}-\kappa _{i_{1}}}%
\right\vert \cdot \left\vert \dfrac{\kappa _{j}+\kappa _{i_{2}}}{\kappa
_{j}-\kappa _{i_{2}}}\right\vert ...\left\vert \dfrac{\kappa _{j}+\kappa
_{i_{\left[ N/2\right] }}}{\kappa _{j}-\kappa _{i_{\left[ N/2\right] }}}%
\right\vert \cosh (\alpha _{i_{1}...i_{\left[ N/2\right] }}+\beta
_{i_{1}...i_{\left[ N/2\right] }}),  \notag
\end{gather}%
and Eq. (\ref{v12}) can be rewritten as%
\begin{equation}
V(x)=-2C\dfrac{d^{2}}{dx^{2}}\left\{ \ln \tau _{N}(x)\right\} .
\label{Alt24}
\end{equation}

In general, as mentioned, the reflectionless potential is uniquely
determined if $2N$ parameters $\kappa _{n}$ and $C_{n}$ ($n=1,2,...,N$) are
known. However, if one sets an additional constraint%
\begin{equation*}
V(-x)=-V(x)
\end{equation*}%
then the number of necessary input parameters is twofold reduced. In other
words, a symmetric reflectionless potential is uniquely determined by its $N$
binding energies \cite{Fermi} $E_{n}=-C\kappa _{n}^{2}$. Let us analyze this
in more detail.

Obviously, Eq. (\ref{Alt23}) can only be symmetric if the arguments of all $%
\cosh $ functions are of the linear form $ax+b$ with $b\equiv 0$. It means,
for example, that%
\begin{equation}
\sum_{i=1}^{N}\kappa _{i}\,x_{i}=\beta _{0}=\sum_{1\leq i<j\leq N}c_{ij},
\label{Alt241}
\end{equation}%
\begin{equation}
\kappa _{1}\,x_{1}-\kappa _{2}\,x_{2}-...-\kappa _{N}\,x_{N}=2\kappa
_{1}\,x_{1}-\beta _{0}=-\beta _{1}=\sum_{j=2}^{N}c_{1j}-\beta _{0},
\label{Alt242}
\end{equation}%
where 
\begin{equation}
c_{ij}\equiv \ln \left\vert \frac{\kappa _{j}+\kappa _{i}}{\kappa
_{j}-\kappa _{i}}\right\vert ,  \label{Alt243}
\end{equation}%
and consequently, 
\begin{equation*}
2\kappa _{1}\,x_{1}=\sum_{j=2}^{N}c_{1j}.
\end{equation*}%
Using Eqs. (\ref{v17}) and Eq. (\ref{v23}), we get a similar expression for
any other combination $\kappa _{i}\,x_{i}.$ From Eqs. (\ref{v10}) and (\ref%
{Alt243}) we therefore obtain the following symmetricity conditions:%
\begin{equation}
\exp \left( 2\kappa _{i}\,x_{i}\right) =\dfrac{C_{i}^{2}}{2\kappa _{n}}%
=\prod\limits_{j\neq i}^{N}\left\vert \dfrac{\kappa _{j}+\kappa _{i}}{\kappa
_{j}-\kappa _{i}}\right\vert ,\ i=1,2,...,N\geq 2.  \label{Alt245}
\end{equation}%
If $N=1$ then%
\begin{equation}
\exp \left( 2\kappa _{1}\,x_{1}\right) =\dfrac{C_{1}^{2}}{2\kappa _{1}}=1.
\label{Alt2451}
\end{equation}

It can be easily shown that Eqs. (\ref{Alt245}) are indeed the symmetricity
conditions for $\tau _{N}$ and $V(x)$. To this end, in full analogy with Eq.
(\ref{Alt242}), we can write%
\begin{gather}
\kappa _{1}\,x_{1}-\kappa _{2}\,x_{2}-...-\kappa _{N}\,x_{N}=\sum_{j\neq
1}^{N}c_{1j}-\beta _{0},  \label{Alt246} \\
-\kappa _{1}\,x_{1}+\kappa _{2}\,x_{2}-...-\kappa _{N}\,x_{N}=\sum_{j\neq
2}^{N}c_{2j}-\beta _{0},  \notag \\
...  \notag \\
-\kappa _{1}\,x_{1}+\kappa _{2}\,x_{2}-...+\kappa _{N}\,x_{N}=\sum_{j\neq
N}^{N}c_{Nj}-\beta _{0}.  \notag
\end{gather}%
Summing these equations, we get 
\begin{equation*}
-(N-2)\sum_{i=1}^{N}\kappa _{i}\,x_{i}=-(N-2)\beta _{0},
\end{equation*}%
which coincides with Eq. (\ref{Alt241}).

\textbf{Remark:} If $N=2$ then first two equations of the system (\ref%
{Alt246}) are not linearly independent, since%
\begin{equation*}
\beta _{1}=\beta _{0}-\sum_{j\neq 1}^{N}c_{1j}=\beta _{0}-\sum_{j\neq
2}^{N}c_{2j}=\beta _{2}=0,
\end{equation*}%
so that $\kappa _{1}\,x_{1}=\kappa _{2}\,x_{2}.$ Consequently, in this (and
only in this) special case Eqs. (\ref{Alt241})-(\ref{Alt242}) must be
treated as the actual symmetricity conditions, while Eq. (\ref{Alt245})
still remains valid.

The next step is to complement Eq. (\ref{Alt246}), for example, with another
condition%
\begin{equation}
2\left( \kappa _{1}\,x_{1}+\kappa _{2}\,x_{2}\right) =\sum_{j\neq
1}^{N}c_{1j}+\sum_{j\neq 2}^{N}c_{2j},  \label{Alt247}
\end{equation}%
which is a direct conclusion from Eq. (\ref{Alt245}). Thus%
\begin{gather*}
2\left( \kappa _{1}\,x_{1}+\kappa _{2}\,x_{2}\right) -\sum_{i=1}^{N}\kappa
_{i}\,x_{i}=\kappa _{1}\,x_{1}+\kappa _{2}\,x_{2}-\kappa
_{3}\,x_{3}-...-\kappa _{N}\,x_{N}= \\
\sum_{j\neq 1}^{N}c_{1j}+\sum_{j\neq 2}^{N}c_{2j}-\beta _{0}=\beta _{12},
\end{gather*}%
which means that%
\begin{equation*}
\cosh (\alpha _{12}+\beta _{12})=\cosh \left[ \left( \kappa _{1}\,+\kappa
_{2}\,-\kappa _{3}\,-...-\kappa _{N}\,\right) \,x\right]
\end{equation*}%
is a symmetric function. Analogously, one can prove that any other term $%
\cosh (\alpha _{i_{1}i_{2}...}+\beta _{i_{1}i_{2}...})$ in Eq. (\ref{Alt23})
is a symmetric function as well. This in turn proves that the norming
constants $C_{n}$ of a symmetric reflectionless potential are uniquely
determined by the given binding energies $E_{n}$.

\subsection{Few practical examples}

To illustrate the results, let us take, for example, $N=4.$ Then Eq. (\ref%
{Alt23}) reads%
\begin{gather}
T_{4}=\cosh (\alpha _{0}+\beta _{0})+\left( \frac{\kappa _{2}+\kappa _{1}}{%
\kappa _{2}-\kappa _{1}}\right) \left( \frac{\kappa _{3}+\kappa _{1}}{\kappa
_{3}-\kappa _{1}}\right) \left( \frac{\kappa _{4}+\kappa _{1}}{\kappa
_{4}-\kappa _{1}}\right) \cosh (\alpha _{1}+\beta _{1})+  \notag \\
\left( \frac{\kappa _{2}+\kappa _{1}}{\kappa _{2}-\kappa _{1}}\right) \left( 
\frac{\kappa _{3}+\kappa _{2}}{\kappa _{3}-\kappa _{2}}\right) \left( \frac{%
\kappa _{4}+\kappa _{2}}{\kappa _{4}-\kappa _{2}}\right) \cosh (\alpha
_{2}+\beta _{2})+\left( \frac{\kappa _{3}+\kappa _{1}}{\kappa _{3}-\kappa
_{1}}\right) \left( \frac{\kappa _{3}+\kappa _{2}}{\kappa _{3}-\kappa _{2}}%
\right) \times  \notag \\
\left( \frac{\kappa _{4}+\kappa _{3}}{\kappa _{4}-\kappa _{3}}\right) \cosh
(\alpha _{3}+\beta _{3})+\left( \frac{\kappa _{4}+\kappa _{1}}{\kappa
_{4}-\kappa _{1}}\right) \left( \frac{\kappa _{4}+\kappa _{2}}{\kappa
_{4}-\kappa _{2}}\right) \left( \frac{\kappa _{4}+\kappa _{3}}{\kappa
_{4}-\kappa _{3}}\right) \cosh (\alpha _{4}+\beta _{4})+  \notag \\
\left( \frac{\kappa _{3}+\kappa _{1}}{\kappa _{3}-\kappa _{1}}\right) \left( 
\frac{\kappa _{4}+\kappa _{1}}{\kappa _{4}-\kappa _{1}}\right) \left( \frac{%
\kappa _{3}+\kappa _{2}}{\kappa _{3}-\kappa _{2}}\right) \left( \frac{\kappa
_{4}+\kappa _{2}}{\kappa _{4}-\kappa _{2}}\right) \cosh (\alpha _{12}+\beta
_{12})+  \label{Alt25} \\
\left( \frac{\kappa _{2}+\kappa _{1}}{\kappa _{2}-\kappa _{1}}\right) \left( 
\frac{\kappa _{4}+\kappa _{1}}{\kappa _{4}-\kappa _{1}}\right) \left( \frac{%
\kappa _{3}+\kappa _{2}}{\kappa _{3}-\kappa _{2}}\right) \left( \frac{\kappa
_{4}+\kappa _{3}}{\kappa _{4}-\kappa _{3}}\right) \cosh (\alpha _{13}+\beta
_{13})+  \notag \\
\left( \frac{\kappa _{2}+\kappa _{1}}{\kappa _{2}-\kappa _{1}}\right) \left( 
\frac{\kappa _{3}+\kappa _{1}}{\kappa _{3}-\kappa _{1}}\right) \left( \frac{%
\kappa _{4}+\kappa _{2}}{\kappa _{4}-\kappa _{2}}\right) \left( \frac{\kappa
_{4}+\kappa _{3}}{\kappa _{4}-\kappa _{3}}\right) \cosh (\alpha _{14}+\beta
_{14}),  \notag
\end{gather}%
while the symmetricity conditions, according to Eqs. (\ref{Alt242})-(\ref%
{Alt245}), can be given as%
\begin{eqnarray}
\kappa _{1}x_{1}-\kappa _{2}x_{2}-\kappa _{3}x_{3}-\kappa _{4}x_{4}
&=&-\beta _{1}=\ln \left( \frac{\kappa _{3}-\kappa _{2}}{\kappa _{3}+\kappa
_{2}}\frac{\kappa _{4}-\kappa _{2}}{\kappa _{4}+\kappa _{2}}\frac{\kappa
_{4}-\kappa _{3}}{\kappa _{4}+\kappa _{3}}\right) ,  \label{Alt26} \\
-\kappa _{1}x_{1}+\kappa _{2}x_{2}-\kappa _{3}x_{3}-\kappa _{4}x_{4}
&=&-\beta _{2}=\ln \left( \frac{\kappa _{3}-\kappa _{1}}{\kappa _{3}+\kappa
_{1}}\frac{\kappa _{4}-\kappa _{1}}{\kappa _{4}+\kappa _{1}}\frac{\kappa
_{4}-\kappa _{3}}{\kappa _{4}+\kappa _{3}}\right) ,  \notag \\
-\kappa _{1}x_{1}-\kappa _{2}x_{2}+\kappa _{3}x_{3}-\kappa _{4}x_{4}
&=&-\beta _{3}=\ln \left( \frac{\kappa _{2}-\kappa _{1}}{\kappa _{2}+\kappa
_{1}}\frac{\kappa _{4}-\kappa _{1}}{\kappa _{4}+\kappa _{1}}\frac{\kappa
_{4}-\kappa _{2}}{\kappa _{4}+\kappa _{2}}\right) ,  \notag \\
-\kappa _{1}x_{1}-\kappa _{2}x_{2}-\kappa _{3}x_{3}+\kappa _{4}x_{4}
&=&-\beta _{4}=\ln \left( \frac{\kappa _{2}-\kappa _{1}}{\kappa _{2}+\kappa
_{1}}\frac{\kappa _{3}-\kappa _{1}}{\kappa _{3}+\kappa _{1}}\frac{\kappa
_{3}-\kappa _{2}}{\kappa _{3}+\kappa _{2}}\right) .  \notag
\end{eqnarray}%
Summing the corresponding sides of Eq. (\ref{Alt26}) we get%
\begin{equation*}
\kappa _{1}x_{1}+\kappa _{2}x_{2}+\kappa _{3}x_{3}+\kappa _{4}x_{4}+\beta
_{0}=0,
\end{equation*}%
which means that%
\begin{equation*}
\cosh (\alpha _{0}+\beta _{0})=\cosh \left[ (\kappa _{1}+\kappa _{2}+\kappa
_{3}+\kappa _{4})\,x\right]
\end{equation*}%
is a symmetric function.

Analogously, substructing the sides of the last two equations from the
corresponding sides of the first two equations of (\ref{Alt26}), we get%
\begin{equation*}
\kappa _{1}x_{1}+\kappa _{2}x_{2}-\kappa _{3}x_{3}-\kappa _{4}x_{4}+\beta
_{12}=0,
\end{equation*}%
which means that $\cosh (\alpha _{12}+\beta _{12})$ is a symmetric function.
Continuing in a similar manner, it is easy to be convinced that $\cosh
(\alpha _{13}+\beta _{13})$ and $\cosh (\alpha _{14}+\beta _{14})$ are
symmetric functions as well.

\noindent \textbf{Example 1:} To be more specific, let $x_{0}=1/\kappa _{1}$
be the length unit and $E_{0}=C\kappa _{1}^{2}$, the energy unit (i.e., $%
x_{0}=1$ and $E_{0}=1$). The simplest and the best known symmetric potential
then corresponds to%
\begin{equation*}
\kappa _{n}=n,\ n=1,2,3,4.
\end{equation*}%
Therefore, according to Eq. (\ref{Alt25}),%
\begin{gather}
T_{4}=\cosh \left[ (\kappa _{1}+\kappa _{2}+\kappa _{3}+\kappa _{4})\,x%
\right] +\frac{3\cdot 4\cdot 5}{1\cdot 2\cdot 3}\cosh \left[ (\kappa
_{1}-\kappa _{2}-\kappa _{3}-\kappa _{4})\,x\right] +\frac{3\cdot 5\cdot 6}{%
1\cdot 1\cdot 2}\times  \label{Alt28} \\
\cosh \left[ (\kappa _{2}-\kappa _{1}-\kappa _{3}-\kappa _{4})\,x\right] +%
\frac{4\cdot 5\cdot 7}{2\cdot 1\cdot 1}\cosh \left[ (\kappa _{3}-\kappa
_{1}-\kappa _{2}-\kappa _{4})\,x\right] +\frac{5\cdot 6\cdot 7}{3\cdot
2\cdot 1}\times  \notag \\
\cosh \left[ (\kappa _{4}-\kappa _{1}-\kappa _{2}-\kappa _{3})\,x\right] +%
\frac{4\cdot 5\cdot 5\cdot 6}{2\cdot 3\cdot 1\cdot 2}\cosh \left[ (\kappa
_{1}+\kappa _{2}-\kappa _{3}-\kappa _{4})\,x\right] +\frac{3\cdot 5\cdot
5\cdot 7}{1\cdot 3\cdot 1\cdot 1}\times  \notag \\
\cosh \left[ (\kappa _{1}+\kappa _{3}-\kappa _{2}-\kappa _{4})\,x\right] +%
\frac{3\cdot 4\cdot 6\cdot 7}{1\cdot 2\cdot 2\cdot 1}\cosh \left[ (\kappa
_{1}+\kappa _{4}-\kappa _{2}-\kappa _{3})\,x\right] =  \notag \\
\cosh \left( 10x\right) +10\cosh \left( 8x\right) +45\cosh \left( 6x\right)
+120\cosh \left( 4x\right) +210\cosh \left( 2x\right) +126.  \notag
\end{gather}

At first sight Eq. (\ref{Alt28}) may seem impractical. However, with the help
of standard transformation formulas (obtained from corresponding trigonometric
formulas by replacing $x\rightarrow ix)$:%
\begin{gather*}
\cosh \left( 10x\right) =512\cosh ^{10}(x)-1280\cosh ^{8}(x)+1120\cosh
^{6}(x)-400\cosh ^{4}(x)+50\cosh ^{2}(x)-1, \\
\cosh \left( 8x\right) =128\cosh ^{8}(x)-256\cosh ^{6}(x)+160\cosh
^{4}(x)-32\cosh ^{2}(x)+1, \\
\cosh \left( 6x\right) =32\cosh ^{6}(x)-48\cosh ^{4}(x)+18\cosh ^{2}(x)-1, \\
\cosh \left( 4x\right) =8\cosh ^{4}(x)-8\cosh ^{2}(x)+1,\ \cosh \left(
2x\right) =2\cosh ^{2}(x)-1,
\end{gather*}%
the result is as follows:%
\begin{equation*}
T_{4}=512\cosh ^{10}(x).
\end{equation*}%
Thus%
\begin{equation*}
\left[ \ln T_{4}(x)\right] ^{\prime }=\frac{T_{4}^{\prime }}{T_{4}}=10\tanh
(x)
\end{equation*}%
and%
\begin{equation*}
V(x)=-\frac{20}{\cosh ^{2}(x)}=-\frac{N(N+1)}{\cosh ^{2}(x)},
\end{equation*}%
exactly as needed.

\begin{figure}[tbh]
\includegraphics[width=\textwidth]{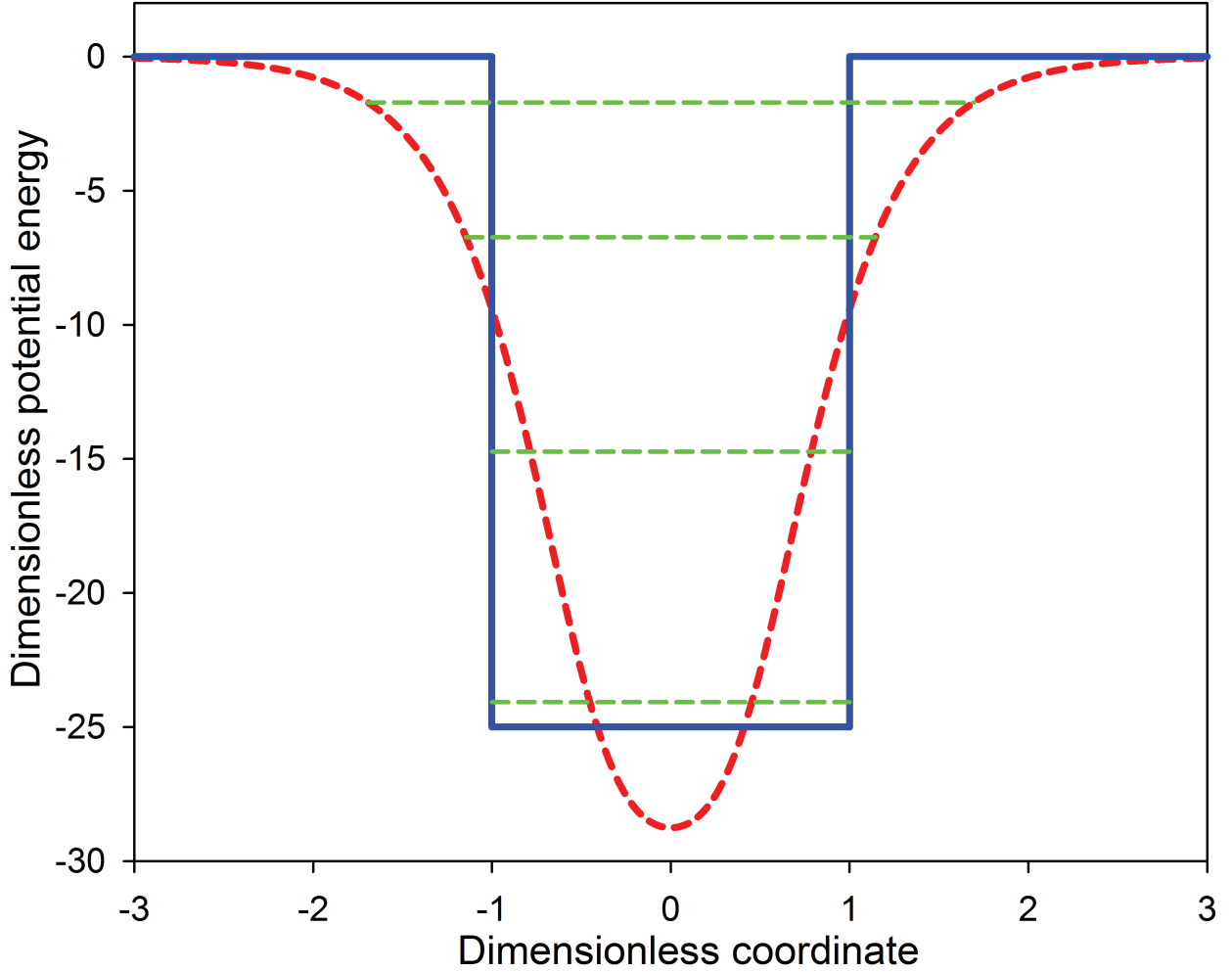} 
\caption{\label{fig:Rect}
The isospectral reflectionless approximant (red dashed line) to a symmetric
rectangular potential (blue solid line). The common energy levels are shown
by green dashed lines.}
\end{figure}

\noindent \textbf{Example 2:} Next, let us construct a reflectionless
approximant to a symmetric rectangular potential with four energy levels
(see Fig. \ref{fig:Rect}). These levels can be determined from (see, e.g., 
\cite{Schiff}, Sect. II.9)%
\begin{eqnarray}
\kappa _{i}\,x_{i}\tan \left( \kappa _{i}\,x_{i}\right) &=&\sqrt{\frac{%
U_{0}a^{2}}{C}-\left( \kappa _{i}\,x_{i}\right) ^{2}},  \label{F1} \\
-\frac{\kappa _{i}\,x_{i}}{\tan \left( \kappa _{i}\,x_{i}\right) } &=&\sqrt{%
\frac{U_{0}a^{2}}{C}-\left( \kappa _{i}\,x_{i}\right) ^{2}},  \label{F2}
\end{eqnarray}%
where $a$ and $U_{0}$ denote the half-width and the depth of the potential
well, respectively. Eq. (\ref{F1}) fixes the symmetric and (\ref{F2}) -- the
antisymmetric solutions to the Schr\"{o}dinger equation. Again, it is
convenient to use dimensionless units for the length and energy, taking $%
x_{0}=a=1$ and $E_{0}=C/a^{2}=1.$ In addition, let us fix%
\begin{equation*}
\sqrt{\frac{U_{0}a^{2}}{C}}=5.
\end{equation*}%
Then the system has four discrete levels (as assumed) corresponding to%
\begin{eqnarray}
\kappa _{1} &=&1.3064400089,  \notag \\
\kappa _{2} &=&2.5957390789,  \label{F3} \\
\kappa _{3} &=&3.8374671080,  \notag \\
\kappa _{4} &=&4.9062951521.  \notag
\end{eqnarray}

In this case Eq. (\ref{Alt25}) cannot be further simplified, but this is not
any serious problem.\ Indeed, let us define the coefficients $A_{i}$ and $%
B_{i}$, such that%
\begin{equation*}
T_{4}=\sum_{i}A_{i}\cosh (B_{i}\,x).
\end{equation*}%
Then the corresponding potential becomes%
\begin{equation}
V(x)=-2C\left\{ \frac{\sum\limits_{i}A_{i}B_{i}^{2}\cosh (B_{i}\,x)}{%
\sum\limits_{i}A_{i}\cosh (B_{i}\,x)}-\left[ \frac{\sum\limits_{i}A_{i}B_{i}%
\sinh (B_{i}\,x)}{\sum\limits_{i}A_{i}\cosh (B_{i}\,x)}\right] ^{2}\right\} .
\label{Alt29}
\end{equation}%
The result for the input data (\ref{F3}) can be seen in Fig. \ref{fig:Rect}.

\begin{figure}[tbh]
\includegraphics[width=\textwidth]{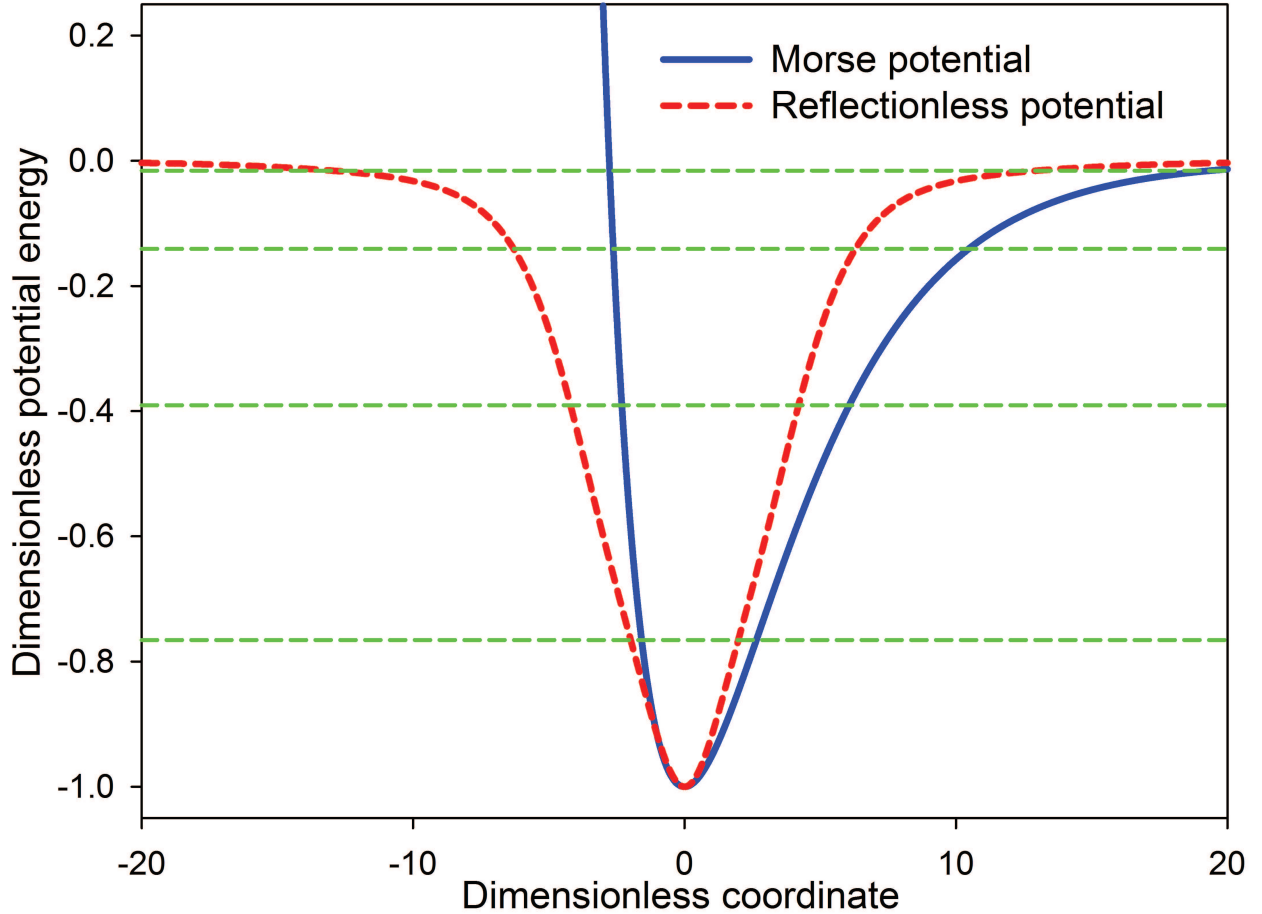} 
\caption{\label{fig:Morse}
The isospectral reflectionless substitute to a Morse potential (see the
explanations to Example 3). The common energy levels are shown by green
dashed lines.}
\end{figure}

\noindent \textbf{Example 3:} Fig. \ref{fig:Morse} demonstrates two
isospectral potentials corresponding to the following set of input
parameters:%
\begin{equation}
\kappa _{1}=1/8,\ \kappa _{2}=3/8,\ \kappa _{3}=5/8,\ \kappa _{4}=7/8,
\label{F35}
\end{equation}%
the four energy levels being $E_{n}=-C\kappa _{n}^{2}$ ($n=1,2,3,4$) as
previously. The solid curve in this figure corresponds to a Morse potential 
\cite{Morse29}%
\begin{equation}
\frac{V(x)}{D}=\exp \left( -\frac{2\alpha x}{x_{0}}\right) -2\exp \left( -%
\frac{\alpha x}{x_{0}}\right) ,  \label{F4}
\end{equation}%
taking $x_{0}\equiv \sqrt{\hbar ^{2}/(2mD)}=1,$ $E_{0}=D=1,$ and
consequently, $C=1.$ The energy eigenvalues read%
\begin{equation}
E_{n}=-D(1-\frac{n+1/2}{a})^{2},  \label{F5}
\end{equation}%
where $a\equiv \sqrt{D/C}/\alpha .$ In Fig. \ref{fig:Morse}, the value $a=4$
has been chosen, so that $\alpha =1/4$ in our dimensionless units. As in the
previous example, the red dashed curve shows the symmetric reflectionless
potential derived by Eq. (\ref{Alt29}) from the input parameters (\ref{F35}).

\section{Conclusion}

The main result of this work is a general formula for calculating $\tau $%
-functions. This important formula, Eq. (\ref{Alt23}), is a direct
conclusion from Theorem 2 that has been proved with the help of well-known
methods of the theory of determinants. We demonstrated that $\tau $%
-functions can be expanded in terms of special determinants called
alternants \cite{Muir}. Any alternant has a divisor -- the Vandermonde's
determinant of the same order, while the quotient can be uniquely expressed
as a polynomial in elementary symmetric functions (\ref{Alt10}) (see Theorem
1). Moreover, in the case of alternants related to the inverse scattering
problem this quotient equals unity, i.e., the alternant itself equals the
Vandermonde's determinant. These useful properties of alternants is the key
to a very simple final result expressed by Eq. (\ref{Alt23}).

Using Eqs. (\ref{Alt23})-(\ref{Alt24}), one can uniquely reconstruct any
reflectionless one-dimensional potential on the full line ($-\infty
<x<\infty $), provided that the $2N$ input parameters $\kappa _{n}$ and $%
C_{n}$ ($n=1,2,...,N$) are known. Moreover, if the result is expected to be
a symmetric function of the coordinate $x$ then the problem can be uniquely
solved on the basis of the $N$ binding energies $E_{n}=-C\kappa _{n}^{2}$.
Compared with the direct use of Eq. (\ref{v9}), the described approach
significantly reduces computational efforts. Indeed, the expansions (%
\ref{v22}) and (\ref{Alt23}) contain only $2^{N-1}$ members, while Eq. (\ref%
{v9}) requires evaluation of a determinant with $N!$ members.

The efficiency of the method has been explicitly demonstrated for the case $%
N=4$, and there is no doubt that the algorithm can be successfully applied
to a much higher number (in principle, to an arbitrary number) of given
binding energies. The described approach can also be applied to building $N$%
-soliton solutions to the Korteweg-de Vries equation, but this would be a
subject for another paper.

\section*{Acknowledgements}

The author acknowledges support from the Estonian Ministry of Education and
Research (target-financed theme IUT2-25) and from ERDF (project
3.2.1101.12-0027) for the research described in this paper.


\begin{thebibliography}{99}
\bibitem{Marchenko1} V. A. Marchenko, \textquotedblleft Some problems of the
theory of differential operators of the second order,\textquotedblright\
Dokl. Akad. Nauk SSSR \textbf{72}, 457-460 (1950).

\bibitem{Marchenko2} V. A. Marchenko, \textquotedblleft Recovery of the
potential energy from the scattering wave phases,\textquotedblright\ Dokl.
Akad. Nauk SSSR \textbf{104}, 695-698 (1955).

\bibitem{GL} I. M. Gel'fand and B. M. Levitan, \textquotedblleft
Determination of a differential equation in terms of its spectral
function,\textquotedblright\ Izv. Akad. Nauk SSSR. Ser. Mat. \textbf{15},
309-360 (1951) [Am. Math. Soc. Transl. (ser. 2) \textbf{1}, 253 (1955)].

\bibitem{Krein1} M. G. Krein, \textquotedblleft On the transfer function of
a one-dimensional boundary value problem of the
second-order,\textquotedblright\ Dokl. Akad. Nauk SSSR \textbf{88}, 405-408
(1953).

\bibitem{Krein2} M. G. Krein, \textquotedblleft On integral equations
generating differential equations of 2nd order,\textquotedblright\ Dokl.
Akad. Nauk SSSR \textbf{97}, 21-24 (1954).

\bibitem{CS} K. Chadan and P. C. Sabatier, \textit{Inverse Problems in
Quantum Scattering Theory}, 2nd ed. (Springer, New York, 1989).

\bibitem{Kay} I. Kay, \textit{The Inverse Scattering Problem} (New York
University, Institute of Mathematical Sciences, Research Report No. EM-74,
1955).

\bibitem{Moses} I. Kay and H. E. Moses, \textquotedblleft The determination
of the scattering potential from the spectral measure function.
III,\textquotedblright\ Nuovo Cimento \textbf{3}, 276-304 (1956).

\bibitem{Faddeev58} L. D. Faddeev, \textquotedblleft Relation of S-matrix
and the potential for the one-dimensional Schr\"{o}dinger
operator,\textquotedblright\ Dokl. Akad. Nauk SSSR \textbf{121}, 63-66
(1958) [Math. Rev. \textbf{20}, 773 (1959)].

\bibitem{Faddeev59} L. D. Faddeev, \textquotedblleft Inverse problem of
quantum scattering theory,\textquotedblright\ Usp. Mat. Nauk \textbf{14},
57-119 (1959) [J. Math. Phys. \textbf{4}, 72 (1963)].

\bibitem{Faddeev64} L. D. Faddeev, \textquotedblleft Properties of the
S-matrix of the one-dimensional Schr\"{o}dinger equation,\textquotedblright\
Trudy Mat. Inst. Akad. Nauk SSSR \textbf{73}, 314 (1964) [Amer. Math. Soc.
Ser. 2 \textbf{65}, 139 (1967)].

\bibitem{Faddeev74} L. D. Faddeev, \textquotedblleft Inverse problem of
quantum scattering theory. II,\textquotedblright\ Itogi Nauki i Tekhniki.
Sovremennye Problemy Matematiki \textbf{3}, 93-180 (1974) [Soviet J. Math. 
\textbf{5} (1976)].

\bibitem{MarchenkoBook} V. A. Marchenko, \textit{Sturm-Liouville Operators
and Applications} (Naukova Dumka, Kiev, 1977) (in Russian) [Birkh\"{a}user,
Basel, 1986; revised ed.: AMS, 2011].

\bibitem{Thacker} H. B. Thacker, C. Quigg, and J. L. Rosner,
\textquotedblleft Inverse problem for quarkonium systems. I. One-dimensional
formalism and methodology,\textquotedblright\ Phys. Rev. D \textbf{18},
274-286 (1978).

\bibitem{Meyer} C. D. Meyer, \textit{Matrix Analysis and Applied Linear
Algebra} (SIAM, Philadelphia, 2000).

\bibitem{Muir} T. Muir, \textit{A Treatise on the Theory of Determinants}
(Macmillan, London, 1882).

\bibitem{Korn} G. A. Korn and T. M. Korn, \textit{Mathematical Handbook for
Scientists and Engineers} (Dover, New York, 2000).

\bibitem{CLO} D. Cox, J. Little, and D. O'Shea, Ideals, \textit{Varieties
and Algorithms}, 2nd ed. (Springer, New York, 1997).

\bibitem{Fermi} J. F. Schonfeld, W. Kwong, and J. L. Rosner,
\textquotedblleft On the convergence of reflectionless approximations to
confining potentials,\textquotedblright\ Ann. Phys. \textbf{128}, 1-28
(1980).

\bibitem{Schiff} L. I. Schiff, \textit{Quantum Mechanics} (McGraw-Hill, New
York, 1949).

\bibitem{Morse29} P. M. Morse, \textquotedblleft Diatomic Molecules
According to the Wave Mechanics. II. Vibrational Levels,\textquotedblright\ {%
Phys. Rev. }\textbf{34}, 57-64 (1929).
\end{thebibliography}
\end{document}